\documentclass[12pt,oneside]{autart}
\usepackage[latin9]{inputenc}
\usepackage{color}
\usepackage{amsmath}
\usepackage{amssymb}
\usepackage{graphicx}

\makeatletter
\newtheorem{notation}[thm]{Notation}
\newtheorem{assumption}[thm]{Assumption}

\@ifundefined{date}{}{\date{}}
\DeclareRobustCommand*{\cal}{\@fontswitch\relax\mathcal}

\makeatother

\begin{document}
\sloppy

\begin{frontmatter}

\title{Multi-sensor State Estimation over Lossy Channels using Coded Measurements}

\author[dl]{Tianju~Sui}\ead{suitj@mail.dlut.edu.cn}, 
\author[guang,argen]{Damian~Marelli\corauthref{cor}}\ead{Damian.Marelli@newcastle.edu.au}, 
\author[dl]{Ximing~Sun}\ead{sunxm@dlut.edu.cn}, 
\author[newcastle,guang]{Minyue~Fu}\ead{minyue.fu@newcastle.edu.au} 

\corauth[cor]{Corresponding author.}

\address[dl]{School of Control Science and Engineering, Dalian University of Technology, Dalian, China.} 
\address[guang]{School of Automation, Guangdong University of Technology, Guangzhou, China.} 
\address[argen]{French Argentine International Center for Information and Systems Sciences, National Scientific and Technical Research Council, Argentina.} 
\address[newcastle]{School of Electrical Engineering and Computer Science, The University of Newcastle, NSW 2308, Australia.} 

\thanks{This work was not presented at any conference.}

\begin{abstract} 
This paper focuses on a networked state estimation problem for a spatially large linear system with a distributed array of sensors, each of which offers partial state measurements, and the transmission is lossy. We propose a measurement coding scheme with two goals. Firstly, it permits adjusting the communication requirements by controlling the dimension of the vector transmitted by each sensor to the central estimator. Secondly, for a given communication requirement, the scheme is optimal, within the family of linear causal coders, in the sense that the weakest channel condition is required to guarantee the stability of the estimator. For this coding scheme, we derive the minimum mean-square error (MMSE) state estimator, and state a necessary and sufficient condition with a trivial gap, for its stability. We also derive a sufficient but easily verifiable stability condition, and quantify the advantage offered by the proposed coding scheme. Finally, simulations results are presented to confirm our claims.
\end{abstract}

\begin{keyword}
Networked state estimation, Sensor fusion, Packet loss, Minimum mean-square error. 
\end{keyword}

\end{frontmatter}

\section{Introduction}

\begin{figure}[h]
\centering{}\includegraphics[width=8.5cm]{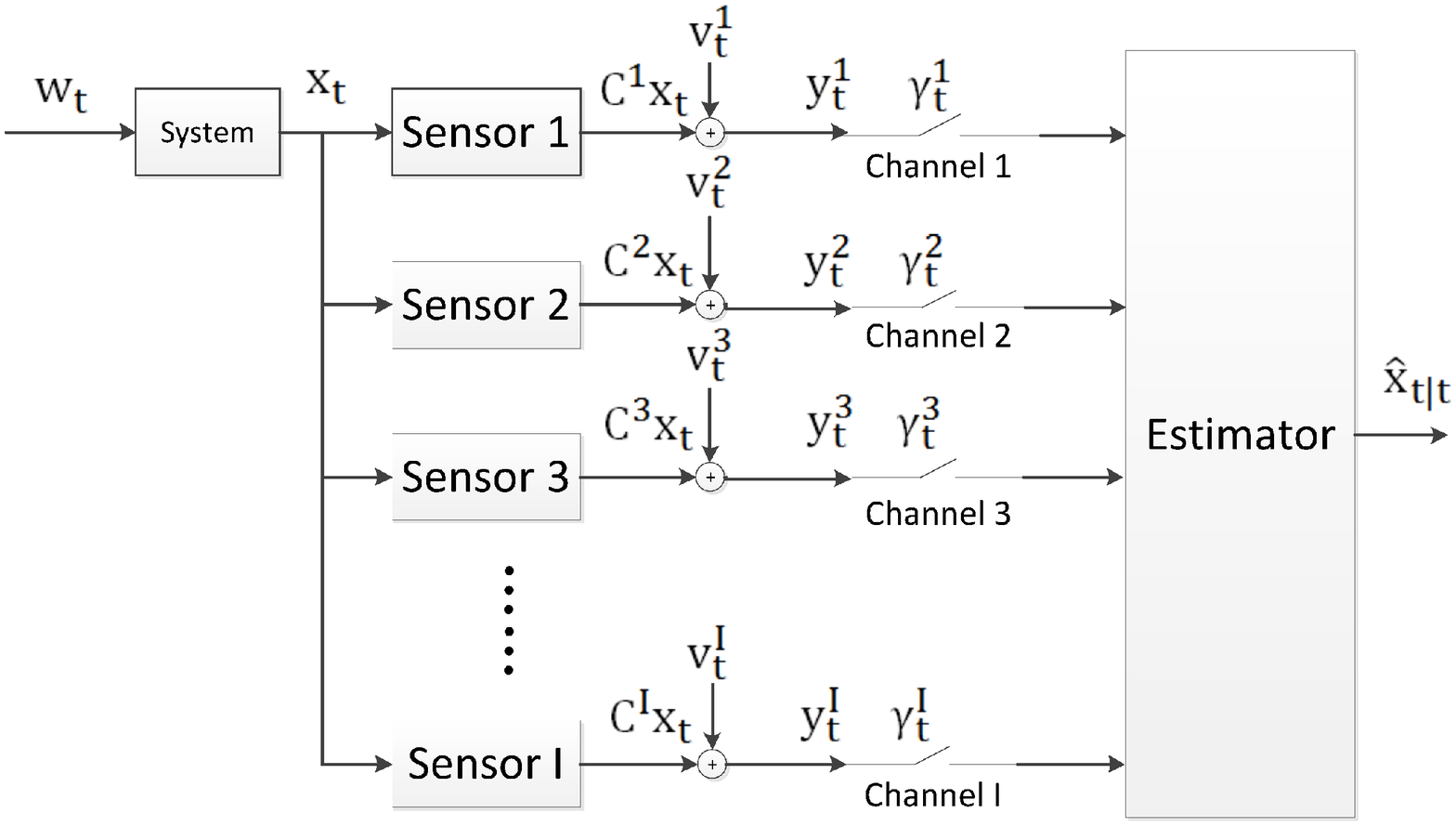} \caption{Networked state estimation using raw measurements.}
\label{fig_conf} 
\end{figure}

This work is concerned with the sensor fusion problem over lossy channels.
Each sensor obtains a partial state measurement subject to some additive
noise, and transmits it to a remote (central) estimator through a
communication network involving packet loss. The estimator computes
a minimum mean-square error (MMSE) estimate of the system state using
the received measurements. The configuration is illustrated in Fig.~\ref{fig_conf}.
This setup is motivated by a wide range of applications including
networked control systems, multi-agent systems, smart electricity
networks and sensor networks~\cite{Gupta2006multi,Liu2004X}.

The problem of networked state estimation, based on MMSE estimation,
has received significant attention in recent years~\cite{sinopoli2004kfi,schenato2007fce,hespanha2007asr,plarre2009kfd}.
One of the major difficulties comes from the packet loss occurring
while transmitting sensor measurements. A central problem consists
in determining the packet loss statistics required to guarantee the
stability of the MMSE estimator. This was done in~\cite{sinopoli2004kfi}
for the case in which the packet loss is independent and identically
distributed. This result has been generalized to different packet
loss models and algebraic system's structure in~\cite{huang2007skf,you2011mean,mo2010towards,mo2008ccv,mo2012kalman,xie2007pcs,rohr2011kalmana,rhor2013general}.
The most general result within this line was recently reported in~\cite{Damian2018Jordan},
where the authors state a conceptional necessary and sufficient condition
with a trivial gap, for general packet loss statistics and system
structure.

The above works assume that raw measurements without preprocessing
are transmitted to the estimator. It turns out that the use of preprocessing
can relax the channel requirements, in terms of channel statistics,
needed to guarantee stability~\cite{Okano2017,Smarra2018}. For example,
in~\cite{schenato2008optimal}, the sensor locally obtains a MMSE
estimate and transmits it instead of its measurement. A drawback of
this approach is that this increases the amount of communications,
because the estimated state needs to be transmitted, which typically
has a higher dimension than the raw measurement. To rectify this,
a \emph{coded measurement}~\cite{Koetter2003,Erez2014} is built
by using a linear combination of the most recent measurements within
a coding window, and this is transmitted instead of the raw measurement~\cite{Lidong2013coding,Sui2014coding}.

The works described so far consider the case in which a single sensor
transmits over a single channel. In many applications, the system
whose state needs to be estimated covers a wide geographical area.
Such a large-scale system is typically equipped with multiple sensors
for measurements. The state estimation problem resulting from this
setup has been studied in a number of works~\cite{he2014networked,hu2013recursive,hu2012extended,wei2009robust,deshmukh2014state,quevedo2013state,Gatsis2015}.
In a sensor network setup, all the sensors can transmit their measurements
to a central estimator over different channels, each with its own
packet loss statistics. Conditions for guaranteeing stability in this
network setup can be very complex, and may be very strong for certain
systems, as reported in~\cite{wei2009robust,deshmukh2014state,quevedo2013state,Sui2015multi}.

In~\cite{Damian2018Jordan}, the authors derived a necessary and
sufficient condition, having a trivial gap, for the stability of a
MMSE estimator. These condition is stated in very general terms, so
it can be applied in a wide range of settings. In the present work,
we make use of this result to design a MMSE estimator for a multi-sensor
network problem. Our contributions are the following: (1) In the context
of this work, the stability of the estimator depends on how reliable
are the communication channels between each sensor and the estimator.
We propose a coding scheme that, while reducing the amount of transmitted
data, i.e., the dimension of the coded vector transmitted by each
sensor at each time step, achieves the weakest requirement on the
channel reliability required to guarantee stability. (2) While the
aforementioned condition is necessary and sufficient, its computation
can be mathematically involved is some cases. To go around this, we
also provide a sufficient condition for easier computation. (3) We
quantify the gain, in terms of channel reliability, offered by the
proposed coding scheme, when compared with the scheme using raw measurements.

The rest of the paper is organized as follows. In Section~\ref{sec:Problem-statement}
we describe the system, channel and coding models. In Section~\ref{sec:estimator},
we derive the expression of the state estimator using coded measurements.
In Section~\ref{subsec:NS} we provide a necessary and sufficient
condition with a trivial gap for the stability of the MMSE estimator.
In Section~\ref{subsec:sufficient} we derive a simpler sufficient
condition for its stability. In Section~\ref{sec:uncoded} we derive
a necessary and sufficient condition with a trivial gap for the stability
of the MMSE estimator using raw measurements, and quantify the advantage
offered by the proposed coding scheme. We give simulation results
illustrating our claims in Section~\ref{sec:simulate}, and give
concluding remarks in Section~\ref{sec:Conclusion}. To improve readability,
some proofs are given in the Appendix.
\begin{notation}
The sets of real and natural numbers are denoted by $\mathbb{R}$
and $\mathbb{N}$, respectively. We use $\mathbb{P}(\mathcal{S})$
to denote the probability of the set $\mathcal{S}$ and $\mathbb{E}(x)$
to denote the expected value of the random variable $x$. For a vector
or matrix $x$ we use $x^{\top}$ to denote its transpose. We use
$I_{d}$ to denote the $d$-dimensional identity matrix and $I$ to
denote the same matrix when the dimension is clear from the context.
\end{notation}

\section{Problem statement\label{sec:Problem-statement}}

Consider a discrete-time stochastic system 
\begin{equation}
x_{t+1}=Ax_{t}+w_{t},\label{sys}
\end{equation}
where $x_{t}\in\mathbb{R}^{n}$ is the system state and $w_{t}\sim\mathcal{N}\left(0,Q\right)$
is white Gaussian noise with $Q\geq0$. The initial time is $t_{0}$
and the initial state is $x_{t_{0}}\sim\mathcal{N}\left(\bar{x}_{t_{0}},P_{t_{0}}\right)$,
with $P_{t_{0}}\geq0$. A sensor network with $I$ nodes, as depicted
in Fig.~\ref{fig_conf}, is used to measure the state in a distributed
manner. For each $i\in\{1,\cdots,I\}$, the measurement $y_{t}^{i}\in\mathbb{R}^{m_{i}}$
obtained at sensor $i$ is given by 
\begin{equation}
y_{t}^{i}=C^{i}x_{t}+v_{t}^{i},\label{eq:msure}
\end{equation}
where $v_{t}^{i}\sim\mathcal{N}\left(0,R^{i}\right)$ is $m_{i}$-dimensional
white Gaussian noise with $R^{i}>0$. Let $m=\sum_{i=1}^{I}m_{i}$
and $C^{\top}=\left[C^{1\top},\cdots,C^{I\top}\right]\in\mathbb{R}^{n\times m}$.
We assume that $(A,C)$ is detectable and $x_{0},w_{t},v_{t}^{i}$
are jointly independent.

We are concerned with a networked estimation system, where each sensor
is linked to the central estimator through a communication network.
Due to the channel unreliability, the transmitted packets may be randomly
lost. We use a binary random process $\gamma_{t}^{i}$ to describe
the packet loss process. That is, $\gamma_{t}^{i}=1$ indicates that
the packet from sensor $i$ is successfully delivered to the estimator
at time $t$, and $\gamma_{t}^{i}=0$ indicates that the packet is
lost. We assume that the random variables $\gamma_{t}^{i}$, $t\in\mathbb{N}$,
$i\in\{1,\cdots,I\}$ are independent and identically distributed
(i.i.d). Also, for each $i\in\left\{ 1,\ldots,I\right\} $, $p_{i}=\mathbb{E}\left[\gamma_{t}^{i}\right]$.

\begin{figure}[h]
\centering{}\includegraphics[width=8.5cm]{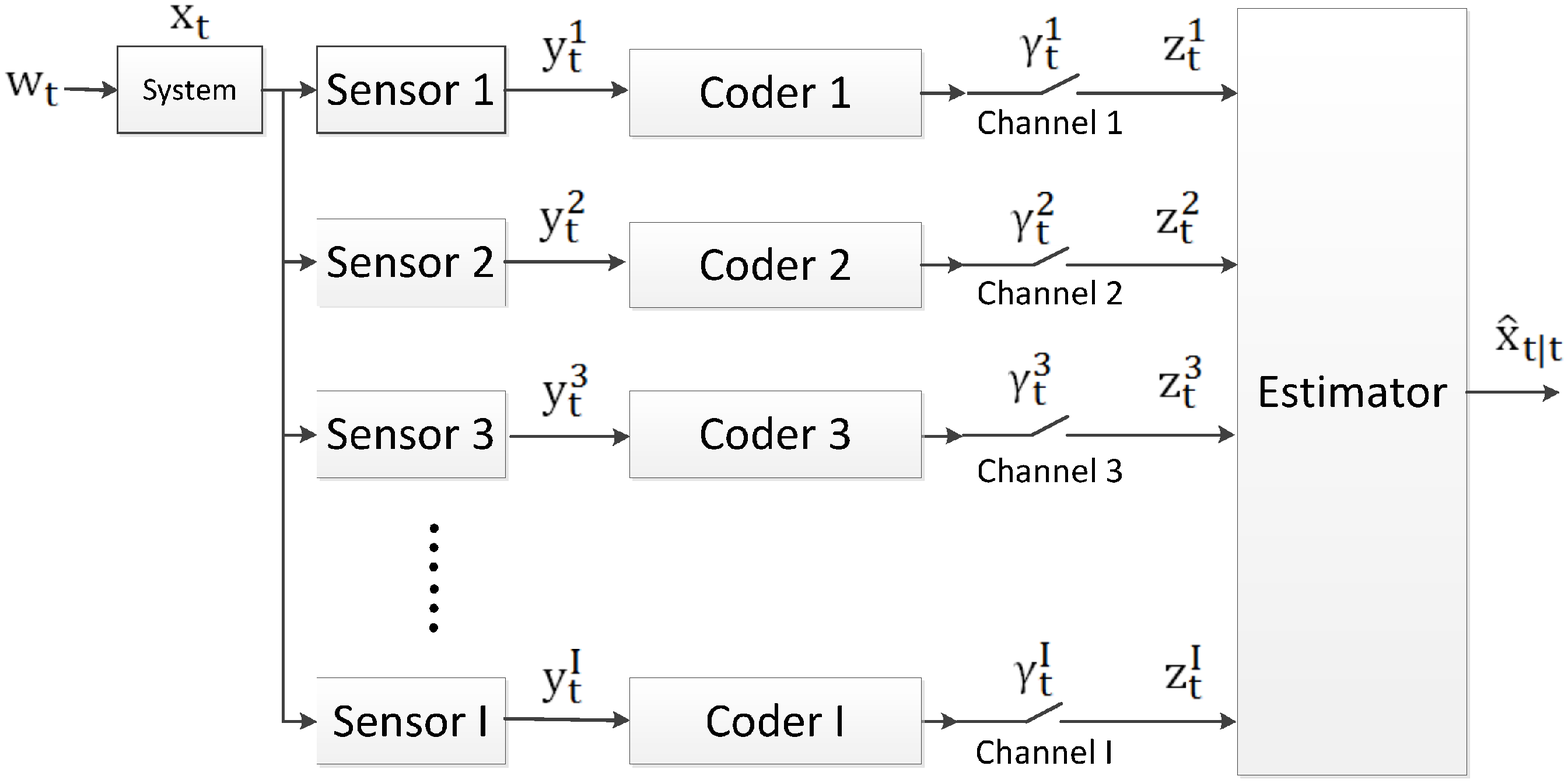}
\caption{Networked state estimation using coded sensor measurements.}
\label{fig_conf2} 
\end{figure}

As a consequence of packet loss, the estimator may fail to generate
a stable state estimate. To improve the stability, instead of transmitting
the raw measurements from each sensor, we encode them before transmission,
as depicted in Fig~\ref{fig_conf2}. More precisely, for a given
\emph{coding window length} $L\in\mathbb{N}$, the coded measurement
of sensor $i$ at time $t$, after going through the channel, is given
by 
\begin{equation}
z_{t}^{i}=\sum_{l=1}^{L}\gamma_{t}^{i}H_{t,l}^{i\top}y_{t-l+1}^{i}\in\mathbb{R}^{c_{i}},\label{eq:coder}
\end{equation}
for some \emph{coding weight} \emph{matrices} $H_{t,l}^{i}\in\mathbb{R}^{c_{i}\times m_{i}}$,
$l\in\{1,\cdots,L\}$, with $c_{i}\le m_{i}$, and the convention
that $y_{t}=0$ for $t\leq t_{0}$.

\begin{rem}
The coding scheme described in~\eqref{eq:coder} allows reducing
the dimension of the transmitted information from $m_{i}$ to $c_{i}$,
to the extent to which even a scalar ($c_{i}=1$) can be transmitted.
This obviously reduce the communication load. We will show in Section~\ref{sec:uncoded}
that the coding scheme can improve the stability of the state estimator
with any choice of $1\le c_{i}\le m_{i}$. 
\end{rem}
To represent the packet loss process for all sensors at time $t$,
we introduce 
\begin{align*}
\Gamma_{t} & =\mathrm{diag}\left\{ \Gamma_{t}^{1},\cdots,\Gamma_{t}^{I}\right\} \in\mathbb{D},\\
\Gamma_{t}^{i} & =\gamma_{t}^{i}I_{c_{i}},
\end{align*}
where $\mathbb{D}$ consists of the $2^{I}$ matrices resulting from
all possible values of $\gamma_{t}^{i}$. The information available
to the estimator from time $t_{0}$ to $t$ is then given by 
\[
\mathcal{F}_{t_{0},t}=\left\{ \left(\Gamma_{t_{0}},\Gamma_{t_{0}}z_{t_{0}}\right),\cdots,\left(\Gamma_{t},\Gamma_{t}z_{t}\right)\right\} ,
\]
where $z_{t}^{\top}=\left[z_{t}^{1\top},\cdots,z_{t}^{I\top}\right]$.
Using this information, the MMSE estimator computes 
\[
\hat{x}_{t|t-1}\left(\mathcal{F}_{t_{0},t}\right)=\mathbb{E}\left[x_{t}|\mathcal{F}_{t_{0},t}\right].
\]
Its prediction error covariance is defined by 
\[
P_{t|t-1}\left(\mathcal{F}_{t_{0},t}\right)=\mathbb{E}\left[\left(x_{t}-\hat{x}_{t|t-1}\right)\left(x_{t}-\hat{x}_{t|t-1}\right)^{\top}|\mathcal{F}_{t_{0},t}\right].
\]

\begin{defn}
We say that the estimator is stable if~\cite{sinopoli2004kfi} 
\[
\sup_{\begin{subarray}{c}
t_{0}\in\mathbb{Z}\\
P_{t_{0}}\geq0
\end{subarray}}\limsup_{t\rightarrow\infty}\left\Vert \mathbb{E}\left(P_{t+1|t}\left(\mathcal{F}_{t_{0},t}\right)\right)\right\Vert <\infty.
\]
\end{defn}
As it is known~\cite{sinopoli2004kfi,schenato2008optimal,rhor2013general},
when the spectral radius of $A$ is greater than one, the packet loss
can lead to an unstable estimator. Our goal is to design $L$ and
$H_{t}$, $t\in\mathbb{N}$, to make the stability condition as weak
as possible. In doing so, we also provide expressions for $\hat{x}_{t|t-1}$
and $P_{t|t-1}$. Notice that, to simplify the notation, we use $\hat{x}_{t|t-1}$
and $P_{t|t-1}$ in place of $\hat{x}_{t|t-1}\left(\mathcal{F}_{t_{0},t}\right)$
and $P_{t|t-1}\left(\mathcal{F}_{t_{0},t}\right)$. We will use this
notation in the rest of the paper.

\section{The MMSE state estimation\label{sec:estimator}}

In this section we assume that $L$ and $H_{t,l}^{i}$, $\forall t,l,i$
are given, and derive the expressions of $\hat{x}_{t|t-1}$ and $P_{t|t-1}$.

Recall that $C^{\top}=\left[C^{1\top},\cdots,C^{I\top}\right]\in\mathbb{R}^{n\times m}$
and define $y_{t}^{\top}=\left[y_{t}^{1\top},\cdots,y_{t}^{I\top}\right]\in\mathbb{R}^{1\times m}$
and $v_{t}^{\top}=\left[v_{t}^{1\top},\cdots,v_{t}^{I\top}\right]\in\mathbb{R}^{1\times m}$.
We can then rewrite~(\ref{eq:msure}) as 
\begin{align*}
y_{t} & =Cx_{t}+v_{t}.
\end{align*}
Let $c=\sum_{i=1}^{I}c_{i}$, $H_{t,l}=\mathrm{diag}\left(\begin{array}{ccc}
H_{t,l}^{1} & \cdots & H_{t,l}^{I}\end{array}\right)\in\mathbb{R}^{c\times m}$ and $H_{t}=\left[H_{t,1},\cdots,H_{t,L}\right]\in\mathbb{R}^{c\times mL}$.
Let also $\Gamma_{t}^{i}=\gamma_{t}^{i}I_{c_{i}}$ and $\Gamma_{t}=\mathrm{diag}\left\{ \Gamma_{t}^{1},\cdots,\Gamma_{t}^{I}\right\} $,
and recall that $z_{t}^{\top}=\left[z_{t}^{1\top},\cdots,z_{t}^{I\top}\right]\in\mathbb{R}^{1\times c}$.
We can then rewrite~(\ref{eq:coder}) as 
\[
z_{t}=\Gamma_{t}H_{t}\left[y_{t}^{\top},\cdots,y_{t-L+1}^{\top}\right]^{\top}.
\]
Let $u_{t}^{\top}=\left[x_{t}^{\top},y_{t}^{\top},\cdots,y_{t-L+1}^{\top}\right]\in\mathbb{R}^{1\times(n+mL)}$.
We can obtain a state-space representation of~(\ref{eq:coder}) as
follows 
\begin{align}
u_{t+1} & =\bar{A}u_{t}+\epsilon_{t},\label{eq:ssx1}\\
z_{t} & =\bar{D}_{t}u_{t},\label{eq:ssx2}
\end{align}
where 
\begin{align*}
\bar{A} & =\left[\begin{array}{ccc}
A & 0 & 0\\
CA & 0 & 0\\
0 & I & 0
\end{array}\right],\quad\epsilon_{t}=\left[\begin{array}{c}
w_{t}\\
Cw_{t}+v_{t+1}\\
0
\end{array}\right],\\
\bar{D}_{t} & =\Gamma_{t}\left[\begin{array}{cc}
0 & H_{t}\end{array}\right]=\left[\bar{D}_{t}^{1},\cdots,\bar{D}_{t}^{I}\right].
\end{align*}

Thus, the expressions of $\hat{x}_{t|t-1}$ and $P_{t|t-1}$ can be
derived by running a Kalman filter\cite{Anderson,sinopoli2004kfi}
on~(\ref{eq:ssx1})-(\ref{eq:ssx2}). The resulting estimator is
given by the following recursions 
\begin{align*}
\hat{x}_{t|t-1} & =\left[I,0\right]\hat{u}_{t|t-1},\\
P_{t|t-1} & =\left[I,0\right]\Sigma_{t|t-1}\left[I,0\right]^{\top},
\end{align*}
where 
\begin{align*}
\hat{u}_{t+1|t} & =\bar{A}\hat{u}_{t|t},\\
\hat{u}_{t|t} & =\hat{u}_{t|t-1}+\sum_{i=1}^{I}\gamma_{k}^{i}K_{t}^{i}\left(z_{t}^{i}-\bar{D}_{t}^{i}\hat{u}_{t|t-1}\right),\\
\Sigma_{t+1|t} & =\bar{A}\Sigma_{t|t}\bar{A}^{\top}+\bar{Q},\\
\Sigma_{t|t} & =\left(I-\sum_{i=1}^{I}\gamma_{k}^{i}K_{t}^{i}\bar{D}_{t}^{i}\right)\Sigma_{t|t-1},
\end{align*}
with 
\begin{align*}
K_{t}^{i} & =\Sigma_{t|t-1}\bar{D}^{i\top}\left(\bar{D}^{i}\Sigma_{t|t-1}\bar{D}^{i\top}\right)^{\dagger},\\
\bar{Q} & =\left[\begin{array}{ccc}
Q & QC^{\top} & 0\\
CQ & CQC^{\top}+R & 0\\
0 & 0 & 0
\end{array}\right],
\end{align*}
and initialized by 
\[
\hat{u}_{t_{0}+1|t_{0}}=0,\quad\Sigma_{t_{0}+1|t_{0}}=\left[\begin{array}{ccc}
P_{t_{0}} & P_{t_{0}}\bar{D}_{t_{0}}^{\top} & 0\\
\bar{D}_{t_{0}}P_{t_{0}} & \bar{D}_{t_{0}}P_{t_{0}}\bar{D}_{t_{0}}^{\top} & 0\\
0 & 0 & 0
\end{array}\right].
\]

\section{Stability Analysis for the MMSE estimator}

In this section we provide conditions for the stability of the MMSE
estimator derived in Section~\ref{sec:estimator}. In Section~\ref{subsec:NS}
we provide a necessary and sufficient condition with a trivial gap,
and give the values of the design parameters $L$ and $H_{t}$, $t\in\mathbb{N}$,
making this condition as weak as possible. In Section~\ref{subsec:sufficient}
we provide a sufficient condition for stability which is easier to
verify.

\subsection{Necessary and sufficient condition\label{subsec:NS}}

In this section we state a necessary and sufficient condition, having
a trivial gap, for the stability of the MMSE estimator using coded
measurements. We then design $L$ and $H_{t}$, $t\in\mathbb{N}$
to make this condition as weak as possible.

Let $\bar{A}=\bar{T}\bar{J}\bar{T}^{-1}$ be the Jordan normal form
of $\bar{A}$. We can then write~(\ref{eq:ssx1})-(\ref{eq:ssx2})
in Jordan canonical form as 
\begin{align}
\tilde{u}_{t+1} & =\bar{J}\tilde{u}_{t}+\tilde{\epsilon}_{t},\label{eq:augSS1}\\
z_{t} & =\tilde{D}_{t}\tilde{u}_{t},\label{eq:augSS2}
\end{align}
where $\tilde{u}_{t}=\bar{T}^{-1}u_{t}$, $\tilde{\epsilon}_{t}=\bar{T}^{-1}\epsilon_{t}$
and $\tilde{D}_{t}=\bar{D}_{t}\bar{T}$. 
\begin{lem}
\label{lem:2} Let $A=TJT^{-1}$ be the Jordan normal form of $A$.
Then 
\begin{equation}
\bar{J}=\mathrm{diag}\left(J,Z_{1},\cdots,Z_{L}\right),\label{eq:Jbar}
\end{equation}
where $Z_{l}$, $l\in\left\{ 1,\cdots,L\right\} $ are $m$-dimensional
Jordan blocks with zero eigenvalues, and there exist matrices $U$
and $V$ such that 
\begin{equation}
\bar{T}=\left[\begin{array}{cc}
T & 0\\
U & V
\end{array}\right].\label{eq:Tbar}
\end{equation}
\end{lem}
\begin{pf}
Let $\bar{t}=\left[0,v^{\top}\right]^{\top}$ with $v^{\top}=\left[v_{1}^{\top},\cdots,v_{L}^{\top}\right]$.
We then have 
\[
\bar{A}\bar{t}=\bar{A}\left[0,v^{\top}\right]^{\top}=\left[0,0,v_{1}^{\top},\cdots,v_{L-1}^{\top}\right]^{\top}.
\]
It follows that $\bar{A}$ has $m$ generalized eigenvectors of rank
$L$ with zero associated eigenvalue. Now, suppose that $t$ is an
eigenvector of $A$ with eigenvalue $\lambda$. Let 
\[
\bar{t}=\left[t^{\top},(Ct)^{\top},\lambda^{-1}(Ct)^{\top},\ldots,\lambda^{-L+2}(Ct)^{\top}\right]^{\top}
\]
It is straightforward to show that $\bar{A}\bar{t}=\lambda\bar{t}$.
Hence, $\bar{t}$ is an eigenvector of $\bar{A}$ with eigenvalue
$\lambda$. Using a similar but somehow more tedious argument, we
can show that, for any generalized eigenvector $t$ of $A$, there
will be a generalized eigenvector $\bar{t}$ of $\bar{A}$, with the
same order and eigenvalue. Hence, the whole set of generalized eigenvectors
of $\bar{A}$ is formed by either $\bar{t}=\left[0,v^{\top}\right]^{\top}$
for some $v$ or $\bar{t}=\left[t^{\top},u^{\top}\right]^{\top}$,
for some $u$ and $t$ being a generalized eigenvector of $A$. Thus,~(\ref{eq:Tbar})
follows. Also, the first $n$ eigenvalues of $\bar{A}$ equal those
of $A$, and the remaining are all zero. Hence,~(\ref{eq:Jbar})
follows. 
\end{pf}
\begin{defn}
A set of complex numbers $x_{i}\in\mathbb{C}$, $i=1,\cdots,I$, is
said to have a common finite multiplicative order $\iota\in\mathbb{N}$
up to a constant $a\in\mathbb{C}$, if $x_{i}^{\iota}=a^{\iota}$,
for all $i=1,\cdots,I$. If there do not exist $\iota$ and $a$ satisfying
the above, the set is said not to have common finite multiplicative
order.\footnote{For example, $e^{\frac{\pi}{2}\jmath}$ and $e^{{\pi}\jmath}$ have
a common finite multiplicative order for that $(e^{\frac{\pi}{2}\jmath})^{4}=(e^{{\pi}\jmath})^{4}=1$.
While, $e^{\frac{\pi}{\sqrt{2}}\jmath}$ and $e^{{\pi}\jmath}$ do
not have a common finite multiplicative order since there does not
exist $k\in\mathbb{N}$ such that $(e^{\frac{\pi}{\sqrt{2}}\jmath})^{k}=(e^{{\pi}\jmath})^{k}=1$.} 
\end{defn}
It is straightforward to see that there is a unique partition of $J$
in diagonal blocks of the form 
\begin{equation}
J=\mathrm{diag}\left(J_{1},\cdots,J_{K}\right),\label{eq:FMO-dec}
\end{equation}
such that, for every $k=1,\ldots,K$, the diagonal entries of the
sub-matrices $J_{k}$ have a common finite multiplicative order $\iota_{k}$
up to $a_{k}$, and for any $k\neq l$, the diagonal entries of the
matrix $\mathrm{diag}(J_{k},J_{l})$ do not have common finite multiplicative
order. Let $d_{k}$ denote the dimension of $J_{k}$ and $a_{k}$
its magnitude. For each $t\in\mathbb{N}$, let 
\[
\tilde{D}_{t}=\left[\tilde{D}_{t,1},\cdots,\tilde{D}_{t,K},\tilde{D}_{t,\ast}\right],
\]
be the partition of $\tilde{D}_{t}$ defined such that, for every
$k=1,\ldots,K$, the number of columns of $\tilde{D}_{t,k}$ equals
the dimension of $J_{k}$. Let 
\[
O_{t,T,k}=\left[\begin{array}{c}
\tilde{D}_{t,k}J_{k}^{t}\\
\vdots\\
\tilde{D}_{t+T-1,k}J_{k}^{t+T-1}
\end{array}\right].
\]

Our next step is to state a necessary and sufficient condition for
the stability of~(\ref{eq:augSS1})-(\ref{eq:augSS2}). To this end,
we aim to use the result in~\cite[Theorem 14]{Damian2018Jordan}.
This result is stated under assumptions which are very general, but
technically involved. Fortunately, we have a way around this technical
difficulty. We have that $\{\tilde{D}_{t}:t\in\mathbb{N}\}$ is a
sequence of random matrices, with discrete distribution, such that
$\{\tilde{D}_{t}:t\in\mathbb{N}\}$ is a statistically independent
set, and whose statistics are cyclostationary with period $M$. Hence,
it follows from~\cite[Proposition 18]{Damian2018Jordan} that the
conditions for~\cite[Theorem 14]{Damian2018Jordan} are guaranteed.
Also, these conditions consider all FMO blocks of~(\ref{eq:augSS1})-(\ref{eq:augSS2}).
In view of Lemma~\ref{lem:2}, this system has $K+1$ blocks. However,
the eigenvalue of the last FMO block equals zero. Hence, the conditions
need only consider the first $K$ blocks. We then obtain the following
result.
\begin{thm}
\label{thm:main} (Combination of~\cite[Proposition 18]{Damian2018Jordan}
and~\cite[Theorem 14]{Damian2018Jordan}) Suppose that the sequence
of coding matrices $\{H_{t}:t\in\mathbb{N}\}$ is $P$-periodic, i.e.,
$H_{t+P}=H_{t}$, for all $t\in\mathbb{N}$. Then, the MMSE estimator
using coded measurements is stable if 
\[
\max_{1\leq k\leq K}|a_{k}|^{2}\Phi_{k}<1,
\]
and unstable if 
\[
\max_{1\leq k\leq K}|a_{k}|^{2}\Phi_{k}>1,
\]
where the \emph{channel unreliability measure $\Phi_{k}$ with respect
to block $k$ is defined by} 
\[
\Phi_{k}=\max_{0\leq t<M}\limsup_{T\rightarrow\infty}\mathbb{P}\left(O_{t,T,k}\text{ does not have FCR}\right)^{1/T},
\]
and $M$ is the least common multiple of $P$ and $\iota_{k}$, $k\in\left\{ 1,\cdots,K\right\} $. 
\end{thm}
\begin{rem}
The result above is inconclusive for the case of $\max_{1\leq k\leq K}|a_{k}|^{2}\Phi_{k}=1$.
For this reason, we say that the necessary and sufficient condition
in Theorem~\ref{thm:main} has a trivial gap. 
\end{rem}
In order to evaluate the condition in Theorem~\ref{thm:main}, we
need to compute the channel unreliability measure $\Phi_{k}$ with
respect to each block $k$. This measure depends on the design parameters
$L$ and $H_{t}$, $t\in\mathbb{N}$. Our next goal is to provide
an expression of $\Phi_{k}$, together with the choices of $L$ and
$H_{t}$, $t\in\mathbb{N}$, so that we can minimize $\Phi_{k}$.

The measure $\Phi_{k}$ is defined in terms of the probability that
the matrix $O_{t,T,k}$ does not have full column rank (FCR). Our
first step towards the computation of $\Phi_{k}$ is to replace $O_{t,T,k}$
by a different matrix, which we denote by $\tilde{O}_{t,T,k}$, for
which the aforementioned probability is easier to compute.

For each $i\in\left\{ 1,\cdots,I\right\} $, let $\tilde{C}^{i}=C^{i}T$
and 
\begin{equation}
\tilde{C}^{i}=\left[\tilde{C}_{1}^{i},\cdots,\tilde{C}_{K}^{i}\right],\label{eq:C^i_k}
\end{equation}
be the partition of $\tilde{C}^{i}$ defined such that, for every
$k=1,\ldots,K$, the number of columns of $\tilde{C}_{k}^{i}$ equals
the dimension of $J_{k}$. Let $H_{t}^{i}=\left[H_{t,1}^{i},\cdots,H_{t,L}^{i}\right]$
and define, for each $k\in\left\{ 1,\cdots,K\right\} $ and $t,T\in\mathbb{N}$,
the following matrix 
\begin{align*}
\tilde{O}_{t,T,k} & =\left[\begin{array}{c}
\tilde{O}_{t,T,k}^{1}\\
\vdots\\
\tilde{O}_{t,T,k}^{I}
\end{array}\right],\qquad\text{with}\qquad\tilde{O}_{t,T,k}^{i}=\left[\begin{array}{c}
\tilde{o}_{k,t}^{i}\\
\vdots\\
\tilde{o}_{k,t+T-1}^{i}
\end{array}\right],\\
\tilde{o}_{k,t}^{i} & =\Gamma_{t}^{i}H_{t}^{i}G_{k}^{i}J_{k}^{t},\qquad\text{and}\qquad G_{k}^{i}=\left[\begin{array}{c}
\tilde{C}_{k}^{i}\\
\vdots\\
\tilde{C}_{k}^{i}J_{k}^{-L+1}
\end{array}\right].
\end{align*}

We have the following result. 
\begin{lem}
\label{lem:Phi_eq} For each $k\in\{1,\cdots,K\}$,

\[
\Phi_{k}=\max_{0\leq t<M}\limsup_{T\rightarrow\infty}\mathbb{P}\left(\tilde{O}_{t,T,k}\text{ does not have FCR}\right)^{1/T}.
\]
\end{lem}
\begin{pf}
See Appendix~\ref{app:Proof-of-NS}. 
\end{pf}
Our next step is to use Lemma~\ref{lem:Phi_eq} to provide an expression
for $\Phi_{k}$, together with the choices of $L$ and $H_{t}$, $t\in\mathbb{N}$,
minimizing its value.

Recall that the packet arrival at time $t$ is represented by the
diagonal matrix $\Gamma_{t}$. Let $\mathbb{D}$ denote the set of
all possible values of $\Gamma_{t}$. We use $\Gamma_{t,T}=\left\{ \Gamma_{t},\cdots,\Gamma_{t+T-1}\right\} \in\mathbb{D}^{T}$
to represent the packet arrivals in the past-time horizon of length
$T$ starting from $t$. For given packet-arrival pattern $S\in\mathbb{D}^{M}$,
we use $\nu^{i}(S)$ to denote the number of measurements from node
$i$ included in $S$.

Let $\mathcal{G}_{k}^{i}$ denote the row span of 
\begin{equation}
\Omega_{k}^{i}=\left[\begin{array}{c}
G_{k}^{i}\\
\vdots\\
G_{k}^{i}J_{k}^{d_{k}-1}
\end{array}\right].\label{eq:regre}
\end{equation}

\begin{defn}
We say that set $\mathcal{L}\subseteq\left\{ 1,\cdots,I\right\} $
of nodes is insufficient for block $k$ if 
\[
\mathrm{span}\left(\bigcup_{i\in\mathcal{L}}\mathcal{G}_{k}^{i}\right)\neq\mathbb{R}^{d_{k}}.
\]
An insufficient set $\mathcal{L}$ is maximal if either $\mathcal{L}=\left\{ 1,\cdots,I\right\} $
or, for all $i\notin\mathcal{L}$, the set $\mathcal{L}\cup\{i\}$
is not insufficient. Let $\mathbb{L}_{k}$ denote the collection of
all maximal insufficient sets for block $k$. 
\end{defn}
Recall that $c_{i}$ is the dimension of $z_{k}^{i}$. For each $\mathcal{L}\in\mathbb{L}_{k}$,
we say that a set of measurement counts $\mathcal{M}=\left\{ 0\leq M^{i}\leq M:i\notin\mathcal{L}\right\} $
is insufficient for $\mathcal{L}$ and $k$ if, for any choice of
$c_{i}M^{i}$ vectors $v_{j}^{i}\in\mathcal{G}_{k}^{i}$, $i\notin\mathcal{L}$,
$j\in\{1,\cdots,c_{i}M^{i}\}$, we have 
\begin{equation}
\mathrm{span}\left(\bigcup_{i\in\mathcal{L}}\mathcal{G}_{k}^{i}\cup\bigcup_{\substack{i\neq\mathcal{L}\\
j\in\{1,\cdots,c_{i}M^{i}\}
}
}v_{j}^{i}\right)\neq\mathbb{R}^{d_{k}}.\label{eq:insuf}
\end{equation}
An insufficient set of counts $\mathcal{M}=\left\{ 0\leq M^{i}\leq M:i\notin\mathcal{L}\right\} $
is maximal if for any $i\notin\mathcal{L}$ for which $M^{i}<M$,
the set obtained by replacing $M^{i}$ by $M^{i}+1$ is not insufficient.
Let $\mathbb{M}_{k}\left(\mathcal{\mathcal{L}}\right)$ denote the
collection of all maximal insufficient sets of counts for $\mathcal{L}$
and $k$. Recall that $p_{i}=\mathbb{E}\left[\gamma_{t}^{i}\right]$,
$i\in\{1,\cdots,I\}$, denotes the packet receival rate for sensor~$i$.
We have the following result. 
\begin{assumption}
\label{assu:pseudo-random}The sequence of coding matrices $\{H_{t}:t\in\mathbb{N}\}$
is $P$-periodic and generated using a pseudo-random sequence with
absolutely continuous distribution. 
\end{assumption}
\begin{thm}
\label{thm:NS} Under Assumption~\ref{assu:pseudo-random}, if $L\geq d_{k}$
for all $k\in\{1,\cdots,K\}$, then w.p.$1$ over the random outcomes
of $H_{t}$, the resulting value of $\Phi_{k}$ is minimized w.r.t.
$L$ and $H_{t}$, $t\in\mathbb{N}$. Furthermore, its value is 
\begin{equation}
\Phi_{k}=\max_{\mathcal{L}\in\mathbb{L}_{k}}\max_{\mathcal{M}\in\mathbb{M}_{k}(\mathcal{L})}\prod_{i\notin\mathcal{L}}\left(1-p_{i}\right)^{1-\frac{M^{i}}{M}},\label{eq:Phi-result}
\end{equation}
where $M$ is the least common multiple of $P$ and $\iota_{k}$ for
any $k\in\left\{ 1,\cdots,K\right\} $.
\end{thm}
\begin{pf}
See Appendix~\ref{app:Proof-of-NS2}. 
\end{pf}
\begin{rem}
We point out that, while Theorem~\ref{thm:NS} asserts that the coding
matrix design the stated in Assumption~\ref{assu:pseudo-random}
is optimal for the purpose of estimator stability, the same design
may not be optimal for the purpose of minimizing the estimation error
covariance. 
\end{rem}
\begin{rem}
For a given FMO block $k\in\{1,\cdots,K\}$, maximal insufficient
set $\mathcal{L}\in\mathbb{L}_{k}$ and node $i\notin\mathcal{L}$,
the vectors $v_{j}^{i}$, $j\in\{1,\cdots,c_{i}M^{i}\}$ belong to
the subspace $\mathcal{G}_{k}^{i}\subset\mathbb{R}^{d_{k}}$. Suppose
that the FMO block $k$ is observable, i.e., $\mathrm{span}\left(\bigcup_{i=1}^{I}\mathcal{G}_{k}^{i}\right)=\mathbb{R}^{d_{k}}$.
Let $\mathcal{M}=\left\{ 0\leq M^{i}\leq M:i\notin\mathcal{L}\right\} \in\mathbb{M}_{k}(\mathcal{L})$
be a maximal insufficient set of counts for $\mathcal{L}$. Then,
since~\eqref{eq:insuf} needs to hold for any choice of $v_{j}^{i}$'s,
it follows that an increment in the dimension $c_{i}$ of the data
transmitted by sensor $i$ would lead to a reduction of the measurement
count $M^{i}$ for that sensor. In view of~\ref{eq:Phi-result},
this will in turn reduce $\Phi_{k}$. Hence, there is a tradeoff between
the communication load (i.e., the value of $c_{i}$ for all $i\in\{1,\cdots,I$\})
and the robustness to packet losses (i.e., the value of $\Phi_{k}$). 
\end{rem}
The above expression of $\Phi_{k}$ greatly simplifies in the limit
case as the period $P$ of the pseudo-random sequence used to generate
$H_{t}$, $t\in\mathbb{N}$ tends to infinity. This is stated in the
following corollary of Theorem~\ref{thm:NS}. This result represents
most practical situations, as periods of pseudo-random sequences are
typically very large. 
\begin{cor}
\label{cor:NS}Under the assumptions of Theorem~\ref{thm:NS}, 
\begin{equation}
\lim_{P\rightarrow\infty}\Phi_{k}=\max_{\mathcal{L}\in\mathbb{L}_{k}}\prod_{i\notin\mathcal{L}}\left(1-p_{i}\right).\label{eq:assymp-cond}
\end{equation}
\end{cor}
\begin{pf}
Notice that, in view of the choices of $L$ and $H_{t}$, if $\mathrm{rank}\left(\tilde{O}_{t,T,k}^{i}\right)<\mathrm{rank}\left(G_{k}^{i}\right),$
then every new measurement from node $i$ yields $\mathrm{rank}\left(\tilde{O}_{t,T+1,k}^{i}\right)\geq\mathrm{rank}\left(\tilde{O}_{t,T,k}^{i}\right)+1$.
Hence, for any $\mathcal{L}\in\mathbb{L}_{k}$ and $\mathcal{M}\in\mathbb{M}_{k}(\mathcal{L})$,
we must have that $M^{i}<\mathrm{rank}\left(G_{k}^{i}\right)$, for
all $i\notin\mathcal{L}$. Since $M$ tends to infinity as so does
$P$, we have 
\begin{align*}
\lim_{P\rightarrow\infty}\Phi_{k} & =\lim_{P\rightarrow\infty}\max_{\mathcal{L}\in\mathbb{L}_{k}}\max_{\mathcal{M}\in\mathbb{M}_{k}(\mathcal{L})}\prod_{i\notin\mathcal{L}}\left(1-p_{i}\right)^{1-\frac{M^{i}}{M}}\\
 & =\max_{\mathcal{L}\in\mathbb{L}_{k}}\max_{\mathcal{M}\in\mathbb{M}_{k}(\mathcal{L})}\prod_{i\notin\mathcal{L}}\left(1-p_{i}\right)\\
 & =\max_{\mathcal{L}\in\mathbb{L}_{k}}\prod_{i\notin\mathcal{L}}\left(1-p_{i}\right).
\end{align*}
\end{pf}
\begin{rem}
The above Corollary~\ref{cor:NS} shows that, when the period $P$
of the pseudo-random sequence used to generate coding matrices is
sufficiently large, the stability condition is no longer affected
by the dimension $c_{i}$ of coded measurements. Nevertheless, a larger
value of $c_{i}$ is still helpful to improve the accuracy of the
estimation. 
\end{rem}

\subsection{An easily verifiable sufficient condition\label{subsec:sufficient}}

The necessary and sufficient condition stated in Theorem~\ref{thm:NS}
requires splitting the system in $K$ blocks. In this section we derive
a condition which is only sufficient, but simpler to compute as it
does not require the aforementioned splitting.

Let 
\begin{align*}
\Xi_{k} & =\left[\begin{array}{c}
\Gamma_{(k-1)P}H_{(k-1)P}F\\
\vdots\\
\Gamma_{kP-1}H_{kP-1}FA^{P-1}
\end{array}\right],
\end{align*}
with 
\[
F=\left[\begin{array}{c}
C\\
\vdots\\
CA^{L-1}
\end{array}\right].
\]

We have the following result. 
\begin{lem}
\label{lem:bounds} If $\Xi_{k}$ does not have full column rank,
\begin{equation}
P_{kP|kP-1}\leq\rho(A)^{2P}\frac{\rho(A)^{2}}{\rho(A)^{2}-1}P_{(k-1)P|(k-1)P-1},\label{eq:Pnot}
\end{equation}
where $\rho\left(A\right)$ denotes the spectral radius of $A$. If
$\Xi_{k}$ has full column rank, there exists $0<\bar{P}\in\mathbb{R}^{n\times n}$
independent of $P_{(k-1)P|(k-1)P-1}$, such that 
\begin{equation}
P_{kP|kP-1}\leq\bar{P}.\label{eq:Pfull}
\end{equation}
\end{lem}
\begin{pf}
See Appendix~\ref{app:Proofs-of-suff}.
\end{pf}
For each $i\in\{1,\cdots,I\}$, let 
\[
F^{i}=\left[\begin{array}{c}
C^{i}\\
\vdots\\
C^{i}A^{L-1}
\end{array}\right].
\]

\begin{defn}
A set $\mathcal{Q}=\left\{ q^{i}\in\mathbb{N}:i=1,\cdots,I\right\} $
of integers is called feasible if, for each $i\in\left\{ 1,\cdots,I\right\} $,
there exist $q^{i}$ indexes $r_{j}^{i}$, $j\in\{1,\cdots,q^{i}\}$,
such that the matrix 
\[
\mathrm{span}\left(\bigcup_{i=1}^{I}\bigcup_{j=1}^{q^{i}}\mathrm{row}_{r_{j}^{i}}\left(F^{i}\right)\right)=\mathbb{R}^{n},
\]
where $\mathrm{row}_{r}\left(X\right)$ denotes the vector formed
by the $r$-th row of matrix $X$. We use $\mathbb{Q}$ to denote
the collection of all feasible sets. 
\end{defn}
We now state the main result of this subsection. 
\begin{thm}
\label{thm:suff}Under Assumption~\ref{assu:pseudo-random}, if $P,L\geq n$,
then, w.p.1 over the random outcomes of $H_{t}$, the MMSE estimator
using coded measurements is stable if
\begin{equation}
\left|\rho(A)\right|^{2}\pi_{P}^{1/P}<1,\label{suff}
\end{equation}
where 
\[
\pi_{P}=\left(\frac{\rho(A)^{2}}{\rho(A)^{2}-1}\right)\sum_{\mathcal{Q}\notin\mathbb{Q}}\prod_{i=1}^{I}\binom{P}{q^{i}}p_{i}^{q_{i}}(1-p_{i})^{P-q_{i}},
\]
with $\binom{P}{q^{i}}$ denoting the binomial coefficient $P$ choose
$q^{i}$. Moreover, if $(A,C^{i})$ is observable for $i=1,\cdots,I$,
then the estimator is unstable if
\begin{equation}
\lim_{P\rightarrow\infty}\left|\rho(A)\right|^{2}\pi_{P}^{1/P}>1.\label{eq:nec}
\end{equation}
\end{thm}
\begin{pf}
See Appendix~\ref{app:Proof-of-NS32}.
\end{pf}

\section{State estimation comparison using raw and coded measurements\label{sec:uncoded}}

In this section we derive the stability condition using raw measurements
to compare with that using coded measurements.

Consider the system described in Section~\ref{sec:Problem-statement}.
Suppose that the raw measurements $y_{t}$, as opposite to the coded
ones $z_{t}$, are transmitted to the estimator, using the same channel
described in Section~\ref{sec:Problem-statement}. The MMSE estimator
then becomes a Kalman filter, having the following information available
at time $t$: 
\[
\breve{\mathcal{F}}_{t}=\left\{ \left(\Gamma_{1},\Gamma_{1}y_{1}\right),\cdots,\left(\Gamma_{t},\Gamma_{t}y_{t}\right)\right\} .
\]
Recall that, for block $k$, $\iota_{k}$ denotes its finite multiplicative
order and $\mathbb{L}_{k}$ denotes the collection of all maximal
insufficient sets. For $\mathcal{L}\in\mathbb{L}_{k}$, we say that
a set $\mathcal{I}(\mathcal{L})=\left\{ 0\leq q_{j}^{i}<\iota_{k}:i\notin\mathcal{L},j=1,\cdots,\check{M}_{i}\right\} $
of indexes is insufficient for block $k$ if 
\[
\breve{\Omega}_{k}=\left[\begin{array}{c}
\breve{\Omega}_{k}^{1}\\
\vdots\\
\breve{\Omega}_{k}^{I}
\end{array}\right]\text{ does not have full column rank},
\]
where, 
\[
\breve{\Omega}_{k}^{i}=\left[\begin{array}{c}
\tilde{C}_{k}^{i}\\
\vdots\\
\tilde{C}_{k}^{i}J_{k}^{\iota_{k}}
\end{array}\right]\text{ if }i\in\mathcal{L}\text{ and }\left[\begin{array}{c}
\tilde{C}_{k}^{i}J_{k}^{q_{1}^{i}}\\
\vdots\\
\tilde{C}_{k}^{i}J_{k}^{q_{\check{M}_{i}}^{i}}
\end{array}\right]\text{ otherwise},
\]
with $\tilde{C}_{k}^{i}$ defined in~(\ref{eq:C^i_k}). We say that
an insufficient set of index is maximal if the set obtained by adding
any extra index is not insufficient. Let $\mathbb{I}_{k}(\mathcal{L})$
denote the collection of maximal insufficient index sets for block
$k$ and set $\mathcal{L}$. For $\mathcal{I}\in\mathbb{I}_{k}(\mathcal{L})$
we use $\nu^{i}(\mathcal{I})$ to denote the number of indexes from
node $i$ included in $\mathcal{I}$.

The following result then states the desired stability condition. 
\begin{prop}
\label{thm:uncoded}The MMSE estimator using raw measurements is stable
if 
\[
\max_{1\leq k\leq K}|a_{k}|^{2}\breve{\Phi}_{k}<1,
\]
and unstable if 
\[
\max_{1\leq k\leq K}|a_{k}|^{2}\breve{\Phi}_{k}>1,
\]
where 
\begin{equation}
\breve{\Phi}_{k}=\max_{\mathcal{L}\in\mathbb{L}_{k}}\max_{\mathcal{I}\in\mathbb{I}_{k}(\mathcal{L})}\prod_{i\notin\mathcal{L}}\left(1-p_{i}\right)^{1-\frac{\nu^{i}(\mathcal{I})}{\iota_{k}}}.\label{eq:raw-cond}
\end{equation}
\end{prop}
\begin{pf}
Using Theorem~\ref{thm:main}, the result holds with 
\[
\breve{\Phi}_{k}=\max_{0\leq t<\iota_{k}}\limsup_{T\rightarrow\infty}\mathbb{P}\left(\breve{O}_{t,T,k}\text{ does not have FCR}\right)^{1/T},
\]
where 
\[
\breve{O}_{t,T,k}=\left[\begin{array}{c}
\left(\Gamma_{t}\otimes I_{m}\right)\tilde{C}_{k}\\
\vdots\\
\left(\Gamma_{T-1}\otimes I_{m}\right)\tilde{C}_{k}J_{k}^{T-1}
\end{array}\right],
\]
with $\tilde{C}_{k}^{\top}=\left[\tilde{C}_{k}^{1\top},\cdots,\tilde{C}_{k}^{I\top}\right]$.
For $S\in\mathbb{D}^{T}$ we use $\breve{O}_{t,T,k}(S)$ to denote
the value of $\breve{O}_{t,T,k}$ resulting when $\Gamma_{t,T}=S$.

Let $\mathbb{S}_{k}\subset\mathbb{D}^{\iota_{k}}$ be the set of all
$S\in\mathbb{D}^{\iota_{k}}$ such that $\breve{O}_{t,\iota_{k},k}(S)$
does not have full column rank. We can then use Lemma~\ref{lem:15}
to obtain 
\[
\breve{\Phi}_{k}=\max_{S\in\mathbb{S}_{k}}\varsigma_{k}(S)^{1/\iota_{k}},
\]
where 
\[
\varsigma_{k}(S)=\mathbb{P}\left\{ \ker\left(\breve{O}_{0,\iota_{k},k}\left(\Gamma_{0,\iota_{k}}\right)\right)\supseteq\ker\left(\breve{O}_{0,\iota_{k},k}\left(S\right)\right)\right\} .
\]
The result then follows after noticing that, if $S_{1}$ contains
the measurements in all entries where $S_{2}$ also does, then $\varsigma_{k}\left(S_{1}\right)\geq\varsigma_{k}\left(S_{2}\right)$.
\end{pf}
We now compare the stability conditions resulting from using coded
and raw measurements. To this end, for the coded case, since the period
$P$ of pseudo-random measurements is typically very large, we use
the asymptotic result given in Corollary~\ref{cor:NS}.

Let $\mathcal{L}_{\mathrm{c}}$ denote the argument which maximizes~(\ref{eq:assymp-cond}).
We have 
\[
\lim_{P\rightarrow\infty}\Phi_{k}=\prod_{i\notin\mathcal{L}_{\mathrm{c}}}\left(1-p_{i}\right),
\]
Let also $\mathcal{L}_{\mathrm{r}}$ be the one maximizing~(\ref{eq:raw-cond}).
We then have 
\begin{align*}
\breve{\Phi}_{k} & =\max_{\mathcal{I}\in\mathbb{I}_{k}\left(\mathcal{L}_{\mathrm{r}}\right)}\prod_{i\notin\mathcal{L}_{\mathrm{r}}}\left(1-p_{i}\right)^{1-\frac{\nu^{i}(\mathcal{I})}{\iota_{k}}}\\
 & \geq\max_{\mathcal{I}\in\mathbb{I}_{k}\left(\mathcal{L}_{\mathrm{c}}\right)}\prod_{i\notin\mathcal{L}_{\mathrm{c}}}\left(1-p_{i}\right)^{1-\frac{\nu^{i}(\mathcal{I})}{\iota_{k}}}\\
 & =\prod_{i\notin\mathcal{L}_{\mathrm{c}}}\left(1-p_{i}\right)\times\max_{\mathcal{I}\in\mathbb{I}_{k}\left(\mathcal{L}_{\mathrm{c}}\right)}\prod_{i\notin\mathcal{L}_{\mathrm{c}}}\left(1-p_{i}\right)^{-\frac{\nu^{i}(\mathcal{I})}{\iota_{k}}}.
\end{align*}
We then obtain 
\begin{align}
\frac{\lim_{P\rightarrow\infty}\Phi_{k}}{\breve{\Phi}_{k}} & \leq\left(\max_{\mathcal{I}\in\mathbb{I}_{k}\left(\mathcal{L}_{\mathrm{c}}\right)}\prod_{i\notin\mathcal{L}_{\mathrm{c}}}\left(1-p_{i}\right)^{-\frac{\nu^{i}(\mathcal{I})}{\iota_{k}}}\right)^{-1}\nonumber \\
 & =\max_{\mathcal{I}\in\mathbb{I}_{k}\left(\mathcal{L}_{\mathrm{c}}\right)}\prod_{i\notin\mathcal{L}_{\mathrm{c}}}\left(1-p_{i}\right)^{\frac{\nu^{i}(\mathcal{I})}{\iota_{k}}}\nonumber \\
 & <1,
\end{align}
which clearly shows the stability improvement offered by the proposed
coding scheme.

\section{Example\label{sec:simulate}}

In this section we use an example to illustrate the improvement, in
terms of the stability of the MMSE estimator, given by the proposed
coding scheme. To this end, we compare the stability of the estimator
in the cases of raw and coded measurements.

We use a system as described in Section~\ref{sec:Problem-statement},
with 
\begin{eqnarray}
A & = & \begin{bmatrix}2 & -4 & -4.5 & 3\\
0 & -2 & -3.5 & 3\\
0 & 0 & 1.5 & 0\\
0 & 0 & 0 & 1
\end{bmatrix}\nonumber \\
 & = & \underbrace{\begin{bmatrix}1 & 1 & 1 & 1\\
0 & 1 & -1 & 1\\
0 & 0 & 1 & 0\\
0 & 0 & 0 & 1
\end{bmatrix}}_{T}\underbrace{\begin{bmatrix}2 & 0 & 0 & 0\\
0 & -2 & 0 & 0\\
0 & 0 & 1.5 & 0\\
0 & 0 & 0 & 1
\end{bmatrix}}_{J}\underbrace{\begin{bmatrix}1 & -1 & -2 & 0\\
0 & 1 & 1 & -1\\
0 & 0 & 1 & 0\\
0 & 0 & 0 & 1
\end{bmatrix}}_{T^{-1}}\label{eq:decomA}
\end{eqnarray}
and $Q=\mathrm{diag}\{1,1,1,1\}$. There are I=4 sensors, with $C^{1}=\begin{bmatrix}1 & 1 & 1 & 1\end{bmatrix}$,
$C^{2}=\begin{bmatrix}1 & -1 & 1 & 1\end{bmatrix}$, $C^{3}=\begin{bmatrix}1 & 1 & 1 & -1\end{bmatrix}$,
$C^{4}=\begin{bmatrix}1 & -1 & 1 & -1\end{bmatrix}$ and $R^{1}=R^{2}=R^{3}=R^{4}=1$.
Also, the communication channels used to transmit sensor measurements,
either raw or coded ones, have packet receival rates $p_{1}=p_{2}=p_{3}=p_{4}=0.4$.

For the coding parameters, we let $L=4$ and $P=500$. For all $t\in\{1,\cdots,P\}$,
$l\in\{1,\cdots,L\}$ and $i\in\{1,\cdots,I\}$, we randomly generate
the entries of $H_{t,l}^{i}$ by drawing them from a standard normal
distribution. Since the period $P=500$ is much larger than the state
dimension $n=4$, we assess the stability of the MMSE estimator using
Corollary~\ref{cor:NS}.

It follows from~\eqref{eq:decomA} that the system is formed by $K=3$
FMO blocks. The magnitude of these blocks are $a_{1}=2$, $a_{2}=1.5$
and $a_{3}=1$, and their finite multiplicative orders are $\iota_{1}=2$,
$\iota_{2}=1$ and $\iota_{3}=1$. Also, the blocks are observable
from all nodes, hence $\mathbb{L}_{1}=\mathbb{L}_{2}=\mathbb{L}_{3}=\emptyset$.
We then have 
\[
\Phi_{1}=\Phi_{2}=\Phi_{3}=\prod_{i=1}^{4}\left(1-p_{i}\right)=\left(1-0.4\right)^{4}=0.1296.
\]
Hence, the filter is stable since 
\[
|a_{3}|^{2}\Phi_{3}<|a_{2}|^{2}\Phi_{2}<|a_{1}|^{2}\Phi_{1}=2^{2}\times0.1296<1.
\]

Since $(A,C^{i})$ is observable, for all $i=1,2,3$, we have that
the sufficient condition in Theorem~\ref{thm:suff} is equivalent
to the necessary and sufficient one given in Theorem~\ref{thm:NS}.
Then, since $P$ is large, we have
\[
(\pi_{P})^{1/P}\simeq0.1296.
\]

For the uncoded case, we start by analyzing the first FMO block. We
have 
\[
\begin{array}{cc}
\tilde{C}_{1}^{1}=\tilde{C}_{1}^{3}=[1,2], & \tilde{C}_{1}^{2}=\tilde{C}_{1}^{4}=[1,0],\\
\tilde{C}_{1}^{1}J_{1}=\tilde{C}_{1}^{3}J_{1}=[2,-4], & \tilde{C}_{1}^{2}J_{1}=\tilde{C}_{1}^{4}J_{1}=[2,0].
\end{array}
\]
Hence, the collection $\mathbb{I}_{1}(\emptyset)=\{\mathcal{I}_{1,1},\mathcal{I}_{1,2},\mathcal{I}_{1,3}\}$
of maximal insufficient index sets contains three sets. Set $\mathcal{I}_{1,1}$
contains index $0$ from node $1$ and index $0$ from node $3$,
the set $\mathcal{I}_{1,2}$ contains index $1$ from node $1$ and
index $1$ from node $3$ and set $\mathcal{I}_{1,3}$ contains indexes
$0$ and $1$ from nodes $2$ and $4$. From Theorem~\ref{thm:uncoded},
we have 
\[
\breve{\Phi}_{1}=\max_{\mathcal{I}\in\mathbb{I}_{1}(\emptyset)}\prod_{i=1}^{4}\left(1-p_{i}\right)^{1-\frac{\nu^{i}(\mathcal{I})}{\iota_{1}}}.
\]
Now, 
\begin{align*}
\prod_{i=1}^{4}\left(1-p_{i}\right)^{1-\frac{\nu^{i}(\mathcal{I}_{1,1})}{\iota_{1}}} & =0.216,\\
\prod_{i=1}^{4}\left(1-p_{i}\right)^{1-\frac{\nu^{i}(\mathcal{I}_{1,2})}{\iota_{2}}} & =0.216,\\
\prod_{i=1}^{4}\left(1-p_{i}\right)^{1-\frac{\nu^{i}(\mathcal{I}_{1,3})}{\iota_{2}}} & =0.36.
\end{align*}
Hence, $\breve{\Phi}_{1}=0.36$, and for the first FMO block we obtain
\[
|a_{1}|^{2}\breve{\Phi}_{1}=2^{2}\times0.36>1.
\]
We therefore do not need to evaluate $\breve{\Phi}_{2}$ and $\breve{\Phi}_{3}$,
since the above inequality in enough to assert that the estimator
is unstable.

\begin{figure}[h]
\centering{}\includegraphics[width=8.5cm]{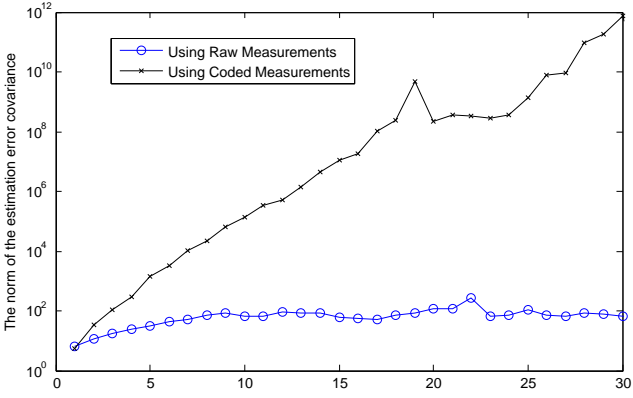} \caption{\textcolor{blue}{Norm of prediction error covariance using coded and
raw measurements.}}
\label{fig:comp} 
\end{figure}

The above claims are illustrated in Figure~\ref{fig:comp}. The figure
shows the time evolution of the norm of the prediction error covariance
yield by the MMSE estimator using both, raw and coded measurements.
To this end we average over $2\times10^{4}$ Monte Carlo runs. The
figure clearly shows that the MMSE estimator is stable when using
coded measurements, while unstable when using raw ones. 

\section{Conclusion\label{sec:Conclusion}}

We studied the networked MMSE state estimation problem for a linear
system with a distributed set of sensors. We proposed a measurement
coding scheme which permits both, controlling the load of communication
used for estimation, and maximizing, within the family of linear causal
coders, the robustness of the resulting estimator against packet losses.
We derived the resulting MMSE estimator, and state a necessary and
sufficient condition, having a trivial gap, for its stability. We
quantified the robustness gain offered by the proposed scheme, by
comparing the stability condition to the one resulting from the use
of raw measurements. We presented simulation results to confirm our
claims.

\appendix

\section{Proofs of Lemma~\ref{lem:Phi_eq}\label{app:Proof-of-NS}}
\begin{notation}
Let $\mathcal{FCR}$ denote the set of matrices having full column
rank. 
\end{notation}
Let 
\[
\bar{T}=\left[\bar{T}_{1},\cdots,\bar{T}_{K},\bar{T}_{\ast}\right],
\]
be the partition of $\bar{T}$ defined such that, for every $k$,
the number of columns of $\bar{T}_{k}$ equals the dimension of $J_{k}$.
We have 
\begin{equation}
\tilde{D}_{t,k}J_{k}^{t}=\bar{D}_{t}\bar{T}_{k}J_{k}^{t}.\label{eq:aux-2}
\end{equation}
Now $\bar{A}\bar{T}=\bar{T}\bar{J}$. Hence, from~(\ref{eq:Jbar}),
\[
\bar{A}\bar{T}_{k}=\bar{T}\mathrm{diag}\left(0,J_{k},0\right)=\bar{T}_{k}J_{k}.
\]
Putting the above into~(\ref{eq:aux-2}) we obtain 
\begin{equation}
\tilde{D}_{t,k}J_{k}^{t}=\bar{D}_{t}\bar{A}^{t}\bar{T}_{k}.\label{eq:aux-1}
\end{equation}

Now, it is straightforward to see that 
\begin{equation}
\bar{A}^{t}=\left[\begin{array}{cc}
A^{t} & 0\\
R_{t} & Q^{t}
\end{array}\right],\quad\text{with}\quad Q=\left[\begin{array}{cc}
0 & 0\\
I & 0
\end{array}\right],\label{eq:Abar^t}
\end{equation}
$R_{0}=0$ and, for $t>0$, 
\begin{equation}
R_{t}=R_{t-1}A+Q^{t-1}R.\label{eq:Rt}
\end{equation}

Putting~(\ref{eq:Abar^t}) into~(\ref{eq:aux-1}), and recalling~(\ref{eq:Tbar}),
we obtain 
\begin{align*}
\tilde{D}_{t,k}J_{k}^{t} & =\Gamma_{t}\left[\begin{array}{cc}
0 & H_{t}\end{array}\right]\left[\begin{array}{cc}
A^{t} & 0\\
R_{t} & Q^{t}
\end{array}\right]\left[\begin{array}{c}
T_{k}\\
U_{k}
\end{array}\right]\\
 & =\Gamma_{t}H_{t}\left[\begin{array}{cc}
R_{t} & Q^{t}\end{array}\right]\left[\begin{array}{c}
T_{k}\\
U_{k}
\end{array}\right]\\
 & =\Gamma_{t}H_{t}R_{t}T_{k}+\Gamma_{t}H_{t}Q^{t}U_{k}.
\end{align*}

Now, for $t\geq L$, we have 
\[
R_{t}=\left[\begin{array}{c}
CA^{t}\\
\vdots\\
CA^{t-L+1}
\end{array}\right].
\]
We then obtain 
\begin{eqnarray*}
 &  & \Gamma_{t}H_{t}R_{t}T_{k}\\
 & = & \Gamma_{t}H_{t}\left[\begin{array}{c}
CA^{t}\\
\vdots\\
CA^{t-L+1}
\end{array}\right]T_{k}=\Gamma_{t}H_{t}\left[\begin{array}{c}
CT_{k}J_{k}^{t}\\
\vdots\\
CT_{k}J_{k}^{t-L+1}
\end{array}\right]\\
 & = & \Gamma_{t}H_{t}\left[\begin{array}{c}
\tilde{C}_{k}\\
\vdots\\
\tilde{C}_{k}J_{k}^{-L+1}
\end{array}\right]J_{k}^{t}=\left[\begin{array}{c}
\tilde{o}_{k,t}^{1}\\
\vdots\\
\tilde{o}_{k,t}^{I}
\end{array}\right]\\
 & \triangleq & \tilde{O}_{t,k}.
\end{eqnarray*}
For $t\geq L$ we also have $Q^{t}=0$, and therefore 
\begin{align*}
\tilde{D}_{t,k}J_{k}^{t} & =\Gamma_{t}H_{t}R_{t}T_{k}\\
 & =\tilde{O}_{t,k}.
\end{align*}
We then have that, for $t\geq L$, $O_{t,T,k}\in\mathcal{FCR}\Leftrightarrow\tilde{O}_{t,T,k}\in\mathcal{FCR}$.
Also, for $t<L$, we obtain 
\begin{align*}
 & ~~~~\limsup_{T\rightarrow\infty}\mathbb{P}\left(O_{t,T,k}\notin\mathcal{FCR}\right)^{1/T}\\
 & =\limsup_{T\rightarrow\infty}\mathbb{P}\left(O_{t,L-t,k}\notin\mathcal{FCR}\right)^{1/T}\mathbb{P}\left(O_{L,T-L,k}\notin\mathcal{FCR}\right)^{1/T}\\
 & =\limsup_{T\rightarrow\infty}\mathbb{P}\left(\tilde{O}_{L,T-L,k}\notin\mathcal{FCR}\right)^{1/T}\\
 & =\limsup_{T\rightarrow\infty}\mathbb{P}\left(\tilde{O}_{t,L-t,k}\notin\mathcal{FCR}\right)^{1/T}\mathbb{P}\left(\tilde{O}_{L,T-L,k}\notin\mathcal{FCR}\right)^{1/T}\\
 & =\limsup_{T\rightarrow\infty}\mathbb{P}\left(\tilde{O}_{t,T,k}\notin\mathcal{FCR}\right)^{1/T},
\end{align*}
and the result follows.

\section{Proofs of Theorem~\ref{thm:NS}\label{app:Proof-of-NS2}}

Let 
\[
E_{t,k}=\Gamma_{t}\left[\begin{array}{c}
H_{t}^{1}G_{k}^{1}\\
\vdots\\
H_{t}^{I}G_{k}^{I}
\end{array}\right].
\]
Clearly, 
\[
\mathrm{rank}\left(\tilde{O}_{t,T,k}\right)=\mathrm{rank}\left(\left[\begin{array}{c}
E_{t,k}J_{k}^{t}\\
\vdots\\
E_{t+T-1,k}J_{k}^{t+T-1}
\end{array}\right]\right).
\]
It then follows from Lemma~\ref{lem:Phi_eq} that, for the purposes
of computing $\Phi_{k}$, the pair $\left(J_{k},\tilde{D}_{t,k}\right)$
is equivalent to $\left(J_{k},E_{t,k}\right)$. For $S\in\mathbb{D}^{M}$
we use $\tilde{O}_{t,M,k}(S)$ to denote the value of $\tilde{O}_{t,M,k}$
resulting when $\Gamma_{t,M}=S$. Let $\mathcal{Z}_{t,k}\subset\mathbb{D}^{M}$
be the set of all $S\in\mathbb{D}^{M}$ such that $\tilde{O}_{t,M,k}(S)\notin\mathcal{FCR}$.

In order to compute $\Phi_{k}$, we make use of the result in~\cite[Proposition 24]{Damian2018Jordan}.
As with~\cite[Theorem 14]{Damian2018Jordan}, this result is stated
under very general assumptions, which are guaranteed by the simpler
assumptions given in~\cite[Proposition 18]{Damian2018Jordan}. Again,
by combining these two results we obtain the following lemma.
\begin{lem}
\label{lem:15} (Combination of~\cite[Proposition 18]{Damian2018Jordan}
and~\cite[Proposition 24]{Damian2018Jordan}) Consider a FMO block
$\left(J_{k},E_{t,k}\right)$. If $E_{t,k}$ is a statistically independent
sequence of random matrices with discrete distribution and cyclostationary
statistics, then
\begin{equation}
\Phi_{k}=\max_{0\leq t<P}\max_{S\in\mathcal{Z}_{t,k}}\varsigma_{t,k}(S)^{1/M}\label{eq:Phi}
\end{equation}
where $M$ is defined in Theorem~\ref{thm:main} and
\[
\varsigma_{t,k}(S)=\mathbb{P}\left\{ \ker\left(\tilde{O}_{t,M,k}\left(\Gamma_{t,M}\right)\right)\supseteq\ker\left(\tilde{O}_{t,M,k}\left(S\right)\right)\right\} .
\]
\end{lem}
Clearly, the pair $\left(J_{k},E_{t,k}\right)$ satisfies the conditions
in Lemma~\ref{lem:15}. Hence we can use the result.

We say that node $i\in\{1,\cdots,I\}$ is complete with respect to
$S\in\mathbb{D}^{M}$ and $k\in\left\{ 1,\cdots,K\right\} $ if $S$
includes $r_{k}=\left\lceil \mathrm{rank}\left(\Omega_{k}^{i}\right)/c_{i}\right\rceil $
measurements from node $i$. Let $\mathcal{C}(S)$ denote the set
of complete nodes in $S$. We have that $\ker\left(\tilde{O}_{t,M,k}\left(\tilde{S}\right)\right)\supseteq\ker\left(\tilde{O}_{t,M,k}\left(S\right)\right)$
if $\tilde{S}$ misses the same measurements on all nodes not in $\mathcal{C}(S)$.
We then have 
\begin{equation}
\varsigma_{t,k}(S)\geq\prod_{i\notin\mathcal{C}(S)}\left(1-p_{i}\right)^{M-\nu^{i}(S)}.\label{eq:geq1}
\end{equation}
Let $\mathcal{N}(S)=\left\{ 0\leq\nu^{i}(S)\leq M:i\notin\mathcal{C}(S)\right\} $
. If $\mathcal{C}(S)$ is an insufficient set and $\mathcal{N}(S)$
are insufficient counts, then $S\in\mathcal{Z}_{t,k}$. Combining
this with~(\ref{eq:Phi}) and~(\ref{eq:geq1}) we obtain 
\begin{align}
\Phi_{k} & \geq\max_{0\leq t<P}\max_{\mathcal{L}\text{ insufficient}}\max_{\mathcal{M}\text{ insufficient}}\prod_{i\notin\mathcal{L}}\left(1-p_{i}\right)^{1-\frac{M^{i}}{M}}\nonumber \\
 & =\max_{\mathcal{L}\in\mathbb{L}_{k}}\max_{\mathcal{M}\in\mathbb{M}_{k}(\mathcal{L})}\prod_{i\notin\mathcal{L}}\left(1-p_{i}\right)^{1-\frac{M^{i}}{M}}.\label{eq:geq2}
\end{align}

Suppose that $L\geq d_{k},k\in\{1,\cdots,K\}$ and $H_{t}$, $t\in\mathbb{N}$,
are generated as in Assumption~\ref{assu:pseudo-random}. Let $S\in\mathbb{D}^{M}$
be such that $\mathcal{C}(S)\in\mathbb{L}_{k}$ and $\mathcal{N}(S)\in\mathbb{M}_{k}(\mathcal{C}(S))$.
Then, w.p.1 over the outcomes of $H_{t}$, any sequence $\tilde{S}$
obtained by adding to $S$ a new measurement from any node $i\notin\mathcal{C}(S)$
will yield $\tilde{O}_{t,M,k}\left(\tilde{S}\right)\in\mathcal{FCR}$.
It then follows that $\ker\left(\tilde{O}_{t,M,k}\left(\tilde{S}\right)\right)\supseteq\ker\left(\tilde{O}_{t,M,k}\left(S\right)\right)$
if and only if $\tilde{S}$ misses the same measurements on all nodes
which are incomplete with respect to $S$. We then have that~(\ref{eq:geq1})
and~(\ref{eq:geq2}) hold with equality, completing the proof.

\section{Proofs of Lemma~\ref{lem:bounds}\label{app:Proofs-of-suff}}

We have 
\begin{eqnarray*}
 &  & P_{kP|kP-1}\\
 & \leq & \mathbb{E}\left[\left(x_{kP}-\hat{x}_{kP|(k-1)P-1}\right)\left(x_{kP}-\hat{x}_{kP|(k-1)P-1}\right)^{\top}\right]\\
 & = & A^{P}P_{(k-1)P|(k-1)P-1}A^{P\top}+\sum_{p=0}^{P-1}A^{p}QA^{p\top}\\
 & \leq & \rho(A)^{2P}P_{(k-1)P|(k-1)P-1}+\sum_{i=0}^{P-1}\rho(A)^{2p}Q\\
 & \leq & \rho(A)^{2P}P_{(k-1)P|(k-1)P-1}+\frac{\rho(A)^{2P}}{\rho(A)^{2}-1}Q,
\end{eqnarray*}
and~(\ref{eq:Pnot}) follows since $P_{(k-1)P|(k-1)P-1}\geq Q$.

Fix $k$ and put $s=(k-1)P-L+1$. For any $(k-1)P\leq t<kP$, it is
straightforward to verify that 
\begin{align}
x_{t} & =A^{t-s}x_{s}+\epsilon_{t-1,s},\label{eq:x_t}\\
y_{t} & =CA^{t-s}x_{s}+\varepsilon_{t,s},\nonumber 
\end{align}
where 
\begin{align*}
\epsilon_{t-1} & =\sum_{r=1}^{t-s}A^{r-1}w_{t-r},\\
\varepsilon_{t,s} & =C\epsilon_{t-1,s}+v_{t}.
\end{align*}
Let $\bar{y}_{t}^{\top}=\left[y_{t-L+1}^{\top},\cdots,y_{t}^{\top}\right]$
and $\bar{\varepsilon}_{t,s}^{\top}=\left[\varepsilon_{t-L+1,s}^{\top},\cdots,\varepsilon_{t,s}^{\top}\right]$.
We then have 
\[
\bar{y}_{t}=FA^{t-(k-1)P}x_{s}+\bar{\varepsilon}_{t,s}.
\]
Let $u_{t}=\Gamma_{t}z_{t}$. Then, 
\begin{align*}
u_{t} & =\Gamma_{t}H_{t}\bar{y_{t}}\\
 & =\Gamma_{t}H_{t}FA^{t-(k-1)P}x_{s}+\Gamma_{t}H_{t}\bar{\varepsilon}_{t,s}.
\end{align*}
Hence, letting $U_{k}^{\top}=\left[u_{(k-1)P}^{\top},\cdots,u_{kP-1}^{\top}\right]$,
we obtain 
\begin{equation}
U_{k}=\Xi_{k}x_{s}+m_{t,s},\label{eq:Z_T}
\end{equation}
where 
\begin{align*}
m_{t,s} & =\left[\begin{array}{c}
\Gamma_{(k-1)P}H_{(k-1)P}\bar{\varepsilon}_{(k-1)P,s}\\
\vdots\\
\Gamma_{kP-1}H_{kP-1}\bar{\varepsilon}_{kP-1,s}
\end{array}\right].
\end{align*}

Using~(\ref{eq:x_t}) and~(\ref{eq:Z_T}), we can obtain an estimate
$\check{x}_{kP}$ of $x_{kP}$ as follows 
\begin{align*}
\check{x}_{kP} & =A^{P}\Xi_{k}^{\dagger}U_{k}.
\end{align*}
If $\Xi_{k}$ has full column rank, then 
\begin{align*}
\check{x}_{kP} & =A^{P}\Xi_{k}^{\dagger}\left(\Xi_{k}x_{s}+m_{t,s}\right)\\
 & =A^{P}x_{s}+\Xi_{k}^{\dagger}m_{t,s}.
\end{align*}
Then, using~(\ref{eq:x_t}), 
\begin{align*}
e_{k} & =x_{kP}-\check{x}_{kP}\\
 & =\epsilon_{t-1,s}-m_{t,s}.
\end{align*}
Since $e_{k}$ is only formed by noise terms, it is straightforward
to see that there exists a matrix $\bar{P}$ such that, w.p.1 and
for all $k\in\mathbb{N}$, 
\[
\mathbb{E}\left[e_{k}e_{k}^{\top}\right]\leq\bar{P}.
\]
The result then follows after noticing that, since $\check{x}_{kP}$
is a sub-optimal estimator, 
\begin{align*}
P_{kP|kP-1} & \leq\mathbb{E}\left[e_{k}e_{k}^{\top}\right].
\end{align*}

\section{Proofs of Theorem~\ref{thm:suff}\label{app:Proof-of-NS32}}

Let $q_{k}^{i}$, $k\in\mathbb{N}$, $i\in\{1,\cdots,I\}$ denote
the number of measurements received from sensor $i$ during the time
interval $[(k-1)P,kP]$. Let $\mathcal{Q}_{k}=\left\{ q_{k}^{i}:i=1,\cdots,I\right\} $.
For $0\leq l\leq k$, let $\mathcal{E}_{l,k}$ denote the event in
which $\mathcal{Q}_{l}\in\mathbb{Q}$ and $\mathcal{Q}_{m}\notin\mathbb{Q}$,
for all $m\in\{l+1,\cdots,k\}$. In particular, notice that $\mathcal{E}_{0,k}$
denotes the event in which $\mathcal{Q}_{m}\notin\mathbb{Q}$, for
all intervals up to $[(k-1)P,kP]$.

Since $H_{t}$, $t\in\mathbb{N}$, is pseudo-randomly generated with
period $P\geq n$, and $L\geq n$, it is straightforward to see that,
with probability one over the outcomes of $H_{t}$, the matrix $Q_{l}$
has full column rank if $\mathcal{Q}_{l}$ is feasible. It then follows
from Lemma~\ref{lem:bounds} that

\begin{eqnarray}
\mathbb{E}\left[P_{kP|kP-1}\right] & = & \sum_{l=0}^{k}\mathbb{E}\left[P_{kP|kP-1}|\mathcal{E}_{l,k}\right]\mathbb{P}\left[\mathcal{E}_{l,k}\right]\nonumber \\
 & \leq & \sum_{l=0}^{k}\kappa^{k-l}\mathbb{E}[P_{lP|lP-1}|\mathcal{E}_{l,k}]\mathbb{P}\left[\mathcal{E}_{l,k}\right]\nonumber \\
 & \leq & \sum_{l=0}^{k}\kappa^{k-l}\mathbb{P}\left[\mathcal{E}_{l,k}\right]\bar{P},\label{eq:aux}
\end{eqnarray}
where 
\[
\kappa=\rho(A)^{2P}\frac{\rho(A)^{2}}{\rho(A)^{2}-1}.
\]

Let 
\[
\varpi_{P}=\mathbb{P}\left[\mathcal{Q}_{l}\notin\mathbb{Q}\right].
\]
We have 
\[
\mathbb{P}\left[\mathcal{E}_{l,k}\right]=\varpi_{P}^{k-l}\left(1-\varpi_{P}\right).
\]
By listing all the possibilities of the event $\mathcal{Q}_{l}\notin\mathbb{Q}$,
it follows that 
\[
\mathbb{P}\left[\mathcal{Q}_{l}\notin\mathbb{Q}\right]=\sum_{\mathcal{Q}\notin\mathbb{Q}}\prod_{i=1}^{I}\binom{P}{q^{i}}p_{i}^{q_{i}}(1-p_{i})^{P-q_{i}}.
\]
 Putting the above into~(\ref{eq:aux}) yields 
\[
\mathbb{E}\left[P_{kP|kP-1}\right]<\left(1-\varpi_{P}\right)\bar{P}\sum_{l=0}^{k}\left(\kappa\varpi_{P}\right)^{k-l}.
\]
Hence, $\sup_{k\in\mathbb{N}}\mathbb{E}\left[P_{kP|kP-1}\right]<\infty$
if 
\begin{align*}
1 & \geq\left(\kappa\varpi_{P}\right)^{1/P}\\
 & =\rho(A)^{2}\left(\frac{\rho(A)^{2}}{\rho(A)^{2}-1}\right)^{1/P}\varpi_{P}^{1/P}\\
 & =\rho(A)^{2}\pi_{P}^{1/P},
\end{align*}
and~(\ref{thm:suff}) the result follows.

Suppose that $(A,C^{i})$ is observable for each $i=1,\ldots,I$.
Then, $\mathcal{Q}\notin\mathbb{Q}$ implies that $q^{i}\leq n$ for
all $i=1,2,\ldots,I$. Thus, we have 
\[
\lim_{P\rightarrow\infty}\pi_{P}^{1/P}=\prod_{i=1}^{I}(1-p_{i}).
\]
Recall from Section~\ref{subsec:NS} that, for each $k=1,\ldots,K$,
$\mathbb{L}_{k}$ denotes the collection of all maximal insufficient
sets. Since all $(A,C^{i})$ are observable, we have $\mathbb{L}_{k}=\emptyset$.
Hence, from Corollary~\ref{cor:NS}, we have 
\[
\lim_{P\rightarrow\infty}\Phi_{k}=\prod_{i=1}^{I}(1-p_{i})
\]
for all $k=1,\ldots,K$. Then,~(\ref{eq:nec}) follows from the necessary
condition stated in Theorem~\ref{thm:main}.

\bibliographystyle{unsrt}
\bibliography{mybibf4}

\end{document}